\documentclass[dvipdfmx,aps,prd,preprintnumbers,superscriptaddress,showpacs,nofootinbib]{revtex4}%
\usepackage[pdftex]{graphicx}
\usepackage{bm,latexsym,amsmath,amssymb,amsfonts,mathrsfs}
\newcommand*{\D}{{\rm d}}

\begin{document}

\title{Constraining DHOST theories with linear growth of matter density fluctuations}


\author{Shin'ichi Hirano}
\email[Email:]{s.hirano"at"rikkyo.ac.jp}
\affiliation{Department of Physics, Rikkyo University, Toshima, Tokyo 171-8501, Japan}

\author{Tsutomu Kobayashi}
\email[Email:]{tsutomu"at"rikkyo.ac.jp}
\affiliation{Department of Physics, Rikkyo University, Toshima, Tokyo 171-8501, Japan}

\author{Daisuke Yamauchi}
\email[Email:]{yamauchi"at"jindai.jp}
\affiliation{Faculty of Engineering, Kanagawa University, Kanagawa, 221-8686, Japan}

\author{Shuichiro Yokoyama}
\email[Email:]{shu"at"kmi.nagoya-u.ac.jp}
\affiliation{Kobayashi Masukawa Institute, Nagoya University, Aichi 464-8602, Japan}
\affiliation{Kavli IPMU (WPI), UTIAS, The University of  Tokyo, Kashiwa, Chiba 277-8583, Japan}


\begin{abstract}
We investigate the potential of cosmological observations, such as galaxy surveys, for constraining degenerate higher-order scalar-tensor (DHOST) theories,
focusing in particular on the linear growth of the matter density fluctuations.
We develop a formalism to describe the evolution of the matter density
fluctuations
during the matter dominated era
and in the early stage of the dark energy dominated era
in DHOST theories,
and give an approximate expression for the gravitational growth index
in terms of several parameters characterizing the theory and
the background solution under consideration.
By employing the current observational constraints on the growth index,
we obtain a new constraint on a parameter space of DHOST theories.
Combining our result with other constraints obtained from
the Newtonian stellar structure,
we show that the degeneracy between the effective parameters of DHOST theories
can be broken without using the Hulse-Taylor pulsar constraint.
\end{abstract}

\preprint{RUP-19-4}
\maketitle

\maketitle

\section{Introduction}

In the era of precision
cosmology, one of the most significant problems is the elucidation of the origin of the late-time acceleration of the Universe~\cite{Riess:1998cb, Perlmutter:1998np}.
Modified gravity theories have been widely studied as an alternative to
the most straightforward candidate of this origin, i.e.,
a cosmological constant.
Among them, scalar-tensor theories with higher-order derivatives
are receiving increasing attention.
In general, such theories would have pathological extra ghost degrees of freedom
because
Ostrogradski's theorem~\cite{Ostro1,Woodard:2015zca}
requires that the equations of motion for a metric and
a scalar field should be of second order to avoid such a ghost degree of freedom.
A new wider class of healthy scalar-tensor theories with higher derivatives
has been proposed recently and
is called the degenerate higher-order scalar-tensor (DHOST)
theories~\cite{Langlois:2015cwa,Crisostomi:2016czh,Achour:2016rkg,BenAchour:2016fzp},
in which the equations of motion are of higher order, but
one can reduce them to a second-order system due to the degeneracy (for a review, see~\cite{Langlois:2017mdk,Langlois:2018dxi,Kobayashi:2019hrl}).
DHOST theories include previously known scalar-tensor theories
such as the Horndeski theory~\cite{Horndeski:1974wa,Deffayet:2011gz,Kobayashi:2011nu}, its disformal relatives~\cite{Zumalacarregui:2013pma}, and (a subclass of) the Gleyzes-Langlois-Piazza-Vernizzi (GLPV) theory~\cite{Gleyzes:2014dya,Gleyzes:2014qga}.
Then, it is intriguing to investigate
observational and experimental constraints on these theories.

One of the most stringent constraints on gravity theories
is obtained from the gravitational wave event
GW170817~\cite{TheLIGOScientific:2017qsa} and its optical counterpart
GRB 170817A~\cite{Monitor:2017mdv},
which gave the constraint on
the speed of gravitational waves, $c_{\rm GW}$, as $|c_{\rm GW}/c-1|\lesssim 10^{-15}$ with $c$ being the speed of light (hereafter we use units in which $c=1$).
This observation can be used to rule out scalar-tensor theories which predict a variable gravitational-wave speed at low redshifts~\cite{Lombriser:2015sxa, Lombriser:2016yzn, Creminelli:2017sry, Sakstein:2017xjx, Ezquiaga:2017ekz, Baker:2017hug, Langlois:2017dyl}.
One finds that there still is a broad class of viable scalar-tensor theories.
In particular, a certain subclass of quadratic DHOST theories~\cite{Langlois:2015cwa, Crisostomi:2016czh, Achour:2016rkg} survived
after this event.

Of course, even before GW170817
lots of stringent constraints on local gravity had been obtained, implying
that
gravity must be consistent with general relativity
at least on small scales and in the weak gravity regime.
Therefore, viable scalar-tensor theories are required to have
a mechanism that suppresses the fifth force mediated by the scalar field
on small scales,
and Vainshtein screening is a typical one of such mechanisms
in the Horndeski and related theories.
Interestingly,
DHOST theories generically exhibit Vainshtein screening outside matter,
whereas its partial breaking occurs
inside~\cite{Kobayashi:2014ida,Langlois:2017dyl,Crisostomi:2017lbg,Dima:2017pwp}.
As the gravitational laws inside an
astrophysical body differ from the standard ones,
this phenomenon leads to a modification of its internal structure,
which can be used to constrain DHOST
 theories~\cite{Saito:2015fza,Sakstein:2015aac,Jain:2015edg,Babichev:2016jom, Saltas:2018mxc, Babichev:2018rfj}.
The authors of Ref.~\cite{Dima:2017pwp}
applied this idea to the DHOST theories
satisfying $c_{\rm GW}^2=1$
and obtained constraints on
the parameters which characterize the theories.

In this paper, in addition to the above constraints, we investigate the possibility of
constraining DHOST theories from the current/future precise cosmological observations.
In particular,
we focus on the linear evolution of the matter density fluctuations,
which can be measured by observations of large scale structure.
Measuring the linear growth rate
of large-scale structure, $f(a)$, is known to be a powerful tool to test
modifications of gravity responsible for the present cosmic acceleration.
To compare the observational data with
theoretical predictions, the simplest approach is to introduce an additional parameter
called gravitational growth index, $\gamma$, defined in terms of the linear growth rate and the fraction parameter of non-relativistic matter $\Omega_{\rm m}$
as~\cite{Wang:1998gt}
\begin{align}
    \gamma:=\frac{\D\ln f}{\D\ln\Omega_{\rm m}}.
\end{align}
The purpose of this paper is
to obtain a novel constraint
on DHOST theories with $c_{\rm GW}^2=1$
from the observations of the linear growth rate.
To do so, we develop a formalism to describe
DHOST cosmology during the matter dominated era and the early stage of the dark energy dominated era, and evaluate the growth index at high redshifts.
We expect that the current observations of the growth index yield
new constraints on DHOST theories which are complementary
to the existing bounds.

The paper is organized as follows.
In Sec.~\ref{sec:DHOST},
we derive cosmological background equations
in class I quadratic DHOST theories. Then
we consider linear cosmological perturbations
and derive the evolution equation of the density fluctuations.
In Sec.~\ref{sec:our model},
we introduce our formalism to model DHOST cosmology and evaluate
the growth index as a probe of modifications gravity. We thereby give novel constraints on DHOST theories from current observations.
Finally, we discuss our results and future prospects in Sec.~\ref{sec:discussion}.

\section{DHOST theories: background and perturbation equations}
\label{sec:DHOST}

\subsection{Action}



The action of the quadratic DHOST theories~\cite{Langlois:2015cwa,Crisostomi:2016czh} is given by
\begin{align}
    S = \int\D^{4}x \sqrt{-g} \left[ G_{2}(\phi,X) -G_{3}(\phi,X)\Box\phi + G_{4}(\phi,X)R
         + \sum^{5}_{i=1}a_{i}(\phi,X)L_{i} \right]
         \label{eq:theory},
\end{align}
where we have several functions of
the scalar field $\phi$ and its kinetic term $X:=(-1/2)\phi_{\mu}\phi^{\mu}$.
The Lagrangians $L_i$ are quadratic in the second derivatives of $\phi$ and are given by
\begin{align}
    L_{1} = \phi_{\mu\nu}\phi^{\mu\nu},\quad  L_{2} = (\Box\phi)^{2},\quad
    L_{3} = (\Box\phi)\phi^{\mu}\phi_{\mu\nu}\phi^{\nu},
\quad
    L_{4} = \phi^{\mu}\phi_{\mu\rho}\phi^{\rho\nu}\phi_{\nu},\quad
    L_{5} = (\phi^{\mu}\phi_{\mu\nu}\phi^{\nu})^{2},
\end{align}
where
$\phi_{\mu}:=\nabla_{\mu}\phi$ and $\phi_{\nu\rho}:=\nabla_{\rho}\nabla_{\nu}\phi$.

In order for this higher-derivative theory to be free of Ostrogradsky ghosts,
we must impose the degeneracy conditions
that relate $G_4$ and $a_i$. The quadratic DHOST theories
are classified in several subclasses~\cite{Langlois:2015cwa, Crisostomi:2016czh},
among which we are interested in the so-called class I theories,
because theories in other subclasses exhibit some pathologies
in a cosmological setup~\cite{deRham:2016wji, Langlois:2017mxy}.
(The class I DHOST theories are conformally/disformally related to
the Horndeski theory~\cite{Achour:2016rkg,Crisostomi:2016czh}.)
The class I degeneracy conditions are summarized as
\begin{align}
a_1+a_2=0,
\quad
\beta_2=-6\beta_1^2,\quad
\beta_3=-2\beta_1\left[
2(1+\alpha_{\rm H})+\beta_1(1+\alpha_{\rm T})
\right],\label{deg_conds}
\end{align}
where
\begin{align}
&M^2 =2(G_4+2Xa_1),
\quad
M^2\alpha_{\rm T}=-4Xa_1,
\quad
M^2\alpha_{\rm H}=-4X(G_{4X}+a_1),
\notag \\
&M^2\beta_1=2X(G_{4X}-a_2+Xa_3),
\quad
M^2\beta_2=4X[a_1+a_2-2X(a_3+a_4)+4X^2a_5],
\notag \\
&M^2\beta_3=-8X(G_{4X}+a_1-Xa_4).
\end{align}
Here we write the derivative of a function $f(X)$ with respect to $X$
as $f_X$. We thus have 3 constraints among 6 functions ($G_4$ and $a_i$),
leaving 3 free functions in addition to $G_2$ and $G_3$.

Note that the propagation speed of gravitational waves is given by
$c_{\rm GW}^2=1+\alpha_{\rm T}$.
The gravitational wave event
GW170817~\cite{TheLIGOScientific:2017qsa} and its optical counterpart GRB 170817A~\cite{Monitor:2017mdv} have placed a tight bound
$c_{\rm GW}^2\simeq 1$. We therefore have $\alpha_{\rm T}\simeq 0$,
provided that this constraint is valid
at low energies where dark energy/modified gravity models are used~\cite{deRham:2018red}.
Imposing $\alpha_{\rm T}=0$ amounts to taking 
$a_1=a_2=0$, but for the moment we do not require this.

 \subsection{Background equations in shift-symmetric DHOST theories}

 In the rest of the paper we focus on the shift-symmetric subclass of
 DHOST theories, in which the Lagrangian is invariant under
 a constant shift of the scalar field, namely $\phi\to\phi+\,$const.
This means that the free functions contained in the Lagrangian are dependent only on the scalar field kinetic term $X$.

As a matter component we only consider pressureless dust and
assume that it is minimally coupled to gravity.
For a homogeneous and isotropic background,
$\D s^2=-\D t^2+a^2(t)\delta_{ij}\D x^i\D x^j$, $\phi=\phi(t)$,
with the matter energy density $\rho_{\rm m}$,
the gravitational field equations read
\begin{align}
3M^2H^2&=  \rho_{\rm m}+ \rho_\phi ,\label{eq:Friedmaneq}
\\
-M^2\left(2\dot H+3H^2\right)&= p_\phi,  \label{eq:Raychaudhurieq}
\end{align}
where $H=\dot a/a $ (a dot denotes differentiation with respect to $t$), and
\begin{align}
\rho_\phi &:= \dot \phi {\cal J}-G_2 -M^2H^2\left( 6\beta_1y
-\frac{1}{2}\beta_2y^2\right),\label{eq:background phi}
\\
p_\phi &:= G_2
+2M^2H^2
\left[
(\alpha_{\rm B}+3\beta_1)y-\left(\beta_1+\frac{\beta_2}{4}\right)y^2
\right]
+2M^2\beta_1\frac{\D}{\D t}\left(yH\right),\label{eq:background_p_phi}
\end{align}
with ${\cal J}$ being the shift current defined shortly.
Here we defined $y:=\ddot\phi/(H\dot\phi)$ and
\begin{align}
\alpha_{\rm M} &:=\frac{1}{M^2H}\frac{\D M^2}{\D t},
\\
\alpha_{\rm B} &:=-\frac{\dot\phi XG_{3X}}{M^2H} +\frac{\alpha_{\rm H}}{y}
-\left(3-\alpha_{\rm M}\right)\beta_1
+\frac{\dot\beta_1}{H}+\left(\beta_1+\frac{\beta_2}{2}\right)y
.
\end{align}
The scalar field equation can be written using the shift current as
\begin{align}
\dot{\cal J}+3H{\cal J}=0,    \label{eq:Fieldeq}
\end{align}
where
\begin{align}
\dot\phi {\cal J}&=
2XG_{2X}+
M^2H^2\left[
\frac{3\alpha_{\rm M}}{y}-6\alpha_{\rm B}+6\left( \alpha_{\rm M}\beta_1+\frac{\dot\beta_1}{H}\right)
+6\beta_1y
-\frac{1}{2}\left( \alpha_{\rm M}\beta_2+\frac{\dot\beta_2}{H}\right)y
\right]
\notag \\ &\quad
+6M^2\beta_{1}\dot H -M^2\beta_2\frac{\D}{\D t}\left(yH\right).
\end{align}
Equation~\eqref{eq:Fieldeq} implies that in the expanding Universe
${\cal J}={\rm const}/a^3\to 0$
and hence attractor solutions are characterized by ${\cal J}=0$.

The background equations~(\ref{eq:Friedmaneq}),~(\ref{eq:Raychaudhurieq}),
and~(\ref{eq:Fieldeq}) contain the higher derivatives
$\dddot\phi$, $\ddddot\phi$, and $\ddot H$.
However, the degeneracy conditions~\eqref{deg_conds}
allow us to reduce the system to the second-order one.
It is not so obvious to demonstrate this explicitly,
but one can follow Refs.~\cite{Crisostomi:2017pjs,Crisostomi:2018bsp} to see that
it is indeed possible to do so.

\subsection{Density perturbations}
\label{sec:perturbations}

Let us study matter density fluctuations in the Newtonian gauge.
The perturbed metric in the Newtonian gauge is given by
\begin{align}
 \D s^{2} = -\left[1+2\Phi (t,{\bm x})\right]\D t^{2} + a^{2}(t)\left[1-2\Psi (t,{\bm x})\right]\delta_{ij}\D x^i\D x^j.
\end{align}
We write the perturbation of the scalar field as
\begin{align}
    \phi (t,{\bm x})= \phi(t) + \pi(t,{\bm x}).
\end{align}
It is convenient to introduce a dimensionless variable
$Q:=H\pi/\dot\phi$, and we will use this instead of $\pi$.
The density perturbation is defined by
\begin{align}
    \rho_{\rm m} (t,{\bm x})= \overline{\rho}_{\rm m}(t)[1+\delta(t,{\bm x})].
\end{align}

We study the quasi-static evolution of the perturbations inside the sound horizon scale. The quasi-static approximation
indicates that
$\dot\epsilon\sim H \epsilon \ll \partial\epsilon$,
where $\epsilon$ is any of perturbation variables.
Expanding the action to second order in perturbations under the quasi-static approximation,
we obtain the following effective action:
\begin{align}
    S_{\mathrm{eff}} = \int\D^{4}x\, {\cal L}_{\mathrm{eff}},
\end{align}
with
\begin{align}
    {\cal L}_{\rm eff}
    &= \frac{M^{2}a}{2}\Biggl\{ (c_{1}\Phi +c_{2}\Psi +c_{3}Q)\partial^{2}Q
        +4(1+\alpha_{\rm H})\Psi\partial^{2}\Phi-2(1+\alpha_{\rm T})\Psi\partial^{2}\Psi
\notag\\
    &\quad -\beta_3\Phi\partial^{2}\Phi +\biggl[4\alpha_{\rm H}\frac{\dot\Psi}{H} -2(2\beta_1 +\beta_3) \frac{\dot\Phi}{H} +(4\beta_1 +\beta_3)\frac{\ddot Q}{H^{2}}\biggr]\partial^{2}Q \Biggr\}
     -a^{3}\overline{\rho}_{\mathrm{m}}\Phi\delta,
\end{align}
where
\begin{align}
    c_1 &:= -4\left[\alpha_{\rm B} -\alpha_{\rm H} +\frac{\beta_3}{2}(1+\alpha_{\rm M}) +\frac{\dot\beta_3}{2H}\right],\label{eq:c1}
\\
    c_2 &:= 4\left[\alpha_{\rm H}(1+\alpha_{\rm M}) +\alpha_{\rm M} -\alpha_{\rm T} +\frac{\dot\alpha_{\rm H}}{H}\right],\label{eq:c2}
\\
    c_3 &:= -2\Bigg\{ \left(1 +\alpha_{\rm M} +\frac{\dot H}{H^2}\right)\left(\alpha_{\rm B} -\alpha_{\rm H}\right)
        +\frac{\dot\alpha_{\rm B} -\dot\alpha_{\rm H}}{H} +\frac{3\Omega_{\rm m}}{2} +\frac{\dot H}{H^2} +\alpha_{\rm T} -\alpha_{\rm M}
\notag\\
    &\quad +\left[-2\frac{\dot H}{H^2}\beta_{1}  +\frac{\beta_3}{4}(1 +\alpha_{\rm M}) +\frac{\dot\beta_3}{2H}\right]\left(1 +\alpha_{\rm M} -\frac{\dot H}{H^2}\right) -2\frac{\dot H}{H^2}\frac{\dot\beta_1}{H}
        +\left(\frac{\dot H}{H^2}\right)^2\frac{\beta_3}{2} +\frac{\dot\alpha_{\rm M}}{H}\frac{\beta_3}{4} +\frac{\ddot\beta_3}{4H^2}  \Bigg\},\label{eq:c3}
        \end{align}
and
\begin{align}
\Omega_{\rm m} &:= \frac{\overline\rho_{\rm m}}{3M^2 H^2}.
\end{align}
We have three terms whose coefficients are written solely in
terms of $\beta_1$ and $\beta_3$.
(The latter can be expressed in terms of $\alpha_{\rm H}$, $\alpha_{\rm T}$, and $\beta_1$
using the degeneracy condition given by Eq.~(\ref{deg_conds}).)
These are the new terms in DHOST theories.
The other terms are present in the Horndeski and GLPV theories,
but as $c_1$ and $c_3$ are dependent on $\beta_1$ and $\beta_3$
one can see implicitly the contributions of these parameters characterizing DHOST theories.

The field equations are derived by varying the effective action with respect to
$\Phi$, $\Psi$, and $Q$.
Going to Fourier space, they are given by
\begin{align}
(1+\alpha_{\rm H})\Psi-\frac{\beta_3}{2}\Phi
+b_1Q+\frac{2\beta_1+\beta_3}{2}\frac{\dot Q}{H}+\frac{a^2}{2M^2k^2}\overline\rho_{\rm m}\delta&=0,
\label{lineq1}\\
(1+\alpha_{\rm T})\Psi-(1+\alpha_{\rm H})\Phi+b_2 Q +\alpha_{\rm H}\frac{\dot Q}{H}&=0,\label{lineq2}\\
c_2\Psi+c_1\Phi + b_3Q+4\alpha_{\rm H}\frac{\dot\Psi}{H}-2(2\beta_1+\beta_3)\frac{\dot\Phi}{H}
+b_4\frac{\dot Q}{H}+2(4\beta_1+\beta_3)\frac{\ddot Q}{H^2}&=0,\label{lineq3}
\end{align}
where $k$ denotes a comoving wavenumber in Fourier space
and $\Phi$, $\Psi$, and $Q$ are now understood as the Fourier components.
Here, the coefficients $b_i$ ($i=1,2,3,4$) are defined as
\begin{align}
b_1&:=\frac{c_1}{4}+\frac{1}{2}(1+\alpha_{\rm M})(2\beta_1+\beta_3)+\frac{1}{2}
\frac{\D}{\D t}\left(\frac{2\beta_1+\beta_3}{H}\right),
\\
b_2&:=-\frac{c_2}{4}+(1+\alpha_{\rm M})\alpha_{\rm H}+\frac{\D}{\D t}
\left(\frac{\alpha_{\rm H}}{H}\right),
\\
b_3&:=2c_3+\left[
\left(1+\alpha_{\rm M}-\frac{\dot H}{H^2}\right)(1+\alpha_{\rm M})+\frac{\dot \alpha_{\rm M}}{H}
\right](4\beta_1+\beta_3)
\notag \\ & \quad
+2(1+\alpha_{\rm M})\frac{\D}{\D t}\left(\frac{4\beta_1+\beta_3}{H}\right)
+\frac{\D^2}{\D t^2}\left(\frac{2\beta_1+\beta_3}{H^2}\right),\label{eq:b3}
\\
b_4&:=2\Biggl[\left(1+\alpha_{\rm M}-\frac{\dot H}{H^2}\right)
(4\beta_1+\beta_3)
+
\frac{\D}{\D t}\left(\frac{4\beta_1+\beta_3}{H}\right)\Biggr].
\end{align}
Since matter is assumed to be minimally coupled to gravity,
the fluid equations are the same as the standard ones, and hence under the quasi-static approximation
the matter density fluctuations $\delta(t,{\bm x}) $ and the velocity field $u^{i}(t,{\bm x}) $
obey
\begin{align}
    &\dot\delta + \frac{1}{a}\partial_{i}[(1+\delta)u^{i}] = 0,
    \label{eq:continuity}
\\
    &\dot u^{i} +Hu^{i} +\frac{1}{a}u^{j}\partial_{j}u^{i} =  -\frac{1}{a}\partial^{i}\Phi.
    \label{eq:Euler}
\end{align}
At linear order,
these equations are combined to give
\begin{equation}
    \ddot{\delta} + 2 H \dot{\delta} + {k^2 \over a^2} \Phi = 0,
    \label{eq:lindelta}
\end{equation}
where we moved to Fourier space.
The effects of modified gravity come into play
through the gravitational potential $\Phi$
which is determined by solving Eqs.~(\ref{lineq1})--(\ref{lineq3}).

Let us then solve Eqs.~(\ref{lineq1})--(\ref{lineq3})
to express $\Phi$, $\Psi$, and $Q$ in terms of $\delta$ and its time derivatives.
We will follow the same procedure as that used in \cite{Hirano:2018uar}.
This procedure is feasible thanks to the degeneracy of the theory.
Solving Eqs.~\eqref{lineq1} and \eqref{lineq2} for $\Phi$ and $\Psi$ and substituting these solutions into Eq.~\eqref{lineq3},
one finds that $\ddot Q$ and $\dot Q$ terms are canceled due to the degeneracy, and hence $Q$ can  be expressed in the form
\begin{align}
    -\frac{k^2}{a^2H^2}Q
        &= \kappa_Q\delta +\nu_Q\frac{\dot\delta}{H},
        \label{eq:Q=delta}
\end{align}
where the explicit expressions for the
time-dependent coefficients $\kappa_Q$ and $\nu_Q$ are presented in
Appendix~\ref{sec:munukappa}.
Finally, substituting this
back into Eqs.~\eqref{lineq1} and \eqref{lineq2}, the gravitational potentials $\Phi$ and $\Psi$
can be expressed in terms of $\delta$, $\dot\delta$, and $\ddot\delta$ as
\begin{align}
    -\frac{k^2}{a^2H^2}\Phi
        &= \kappa_\Phi\delta +\nu_\Phi\frac{\dot\delta}{H}
         +\mu_\Phi\frac{\ddot\delta}{H^2},
         \label{eq:Phi=delta}
\\
         -\frac{k^2}{a^2H^2}\Psi
        &= \kappa_\Psi\delta +\nu_\Psi\frac{\dot\delta}{H}
         +\mu_\Psi\frac{\ddot\delta}{H^2}.
    \label{eq:Psi=delta}
\end{align}
The explicit expressions for
the time-dependent coefficients
$\mu_{i}$, $\nu_{i}$, and $\kappa_{i}$ ($i=\Phi,\Psi$)
are also shown in Appendix~\ref{sec:munukappa}.
Within the Horndeski theory we have $\mu_i=\nu_i=0$ and in the GLPV theory we still have $\mu_\Psi=0$.
That is,
$\mu_\Psi$ first appears in DHOST theories beyond GLPV.
Equation~(\ref{eq:Phi=delta}) allows us to eliminate $\Phi$ from
Eq.~(\ref{eq:lindelta}) and we obtain
the closed-form equation for $\delta$ as
\begin{align}
    \ddot{\delta} + (2+\varsigma)H\dot{\delta} -\frac{3}{2}\Omega_{\mathrm{m}}\Xi_{\Phi}H^{2}\delta = 0
    \label{eq:first-order evolution},
\end{align}
where the additional friction
$\varsigma$ and the effective gravitational coupling
(multiplied by $8\pi M^2$)
$\Xi_{\Phi}$ are written in terms of $\mu_\Phi\,,\nu_\Phi$\,, and $\kappa_\Phi$ as
\begin{align}
    \varsigma & =\frac{2 \mu_\Phi - \nu_\Phi}{1 - \mu_\Phi}
    \,,\label{eq:varsigma def}\\
     \Xi_\Phi &= \frac{2}{3\Omega_{\rm m}}
             \frac{\kappa_\Phi}{1 - \mu_\Phi}
    \,.\label{eq:Xi_Phi def}
\end{align}
These two functions characterize modification of gravity.
The evolution equation~\eqref{eq:first-order evolution}
has essentially the same form as that in DHOST theories with
$c_{\rm GW}^2=1$~\cite{Crisostomi:2017pjs}
and in the GLPV theory~\cite{Gleyzes:2014qga, DAmico:2016ntq}.
Whether or not $c_{\rm GW}^2=1$ does not play an important role
in determining the qualitative form of Eq.~\eqref{eq:first-order evolution}.
In the case of the Horndeski theory ($\alpha_{\rm H}=\beta_1=0$),
the additional friction term vanishes, $\varsigma=0$, and
the result of Ref.~\cite{DeFelice:2011hq} is recovered.

Equation~(\ref{eq:first-order evolution}) tells us that, even
in DHOST theories under the quasi-static approximation,
the evolution of the matter density fluctuations
is independent of the wavenumber, so that as usual we can
write the growing solution to Eq.~(\ref{eq:first-order evolution}) as
\begin{align}
    \delta(t, {\bf k}) = D_+(t)\delta_{\rm L}({\bf k}),
\end{align}
where $\delta_{\rm L}({\bf k})$ represents the initial density field.
The effect of the modified evolution of the density perturbations
is thus imprinted in the growth factor, $D_+(t)$.
Introducing the linear growth rate, $f:=\D\ln D_+/\D\ln a$,
the evolution equation can be written as
\begin{align}
    \frac{\D f}{\D\ln{a}} + \left(2 +\varsigma +\frac{\D\ln{H}}{\D\ln{a}}\right)f +f^{2} -\frac{3}{2}\Omega_{\rm m}\Xi_{\Phi} = 0.
    \label{eq:evolution f}
\end{align}
Given the expansion history and the dynamics of the scalar field, one can obtain the evolution of the linear growth rate by solving the above equation.

\section{Modeling DHOST cosmology in the matter dominated era}
\label{sec:our model}

We consider possible cosmological constraints on
DHOST theories from observables
during the matter dominated era and in the early stage of the dark energy
dominated era.
To do so, we assume that
during these stages $y$, $G_2$,
$\alpha_i$ ($i=$H, M, B, T), and $\beta_1$
can be expressed as a series expansion form
in terms of $\varepsilon:=1-\Omega_{\rm m}\,(\ll 1)$ as
\begin{align}
    y&=y_0+{\cal O}(\varepsilon),
    \label{eq:y0 exp}\\
    G_2&=g_2M^2H^2\varepsilon+{\cal O}\left(\varepsilon^2\right),
    \\
    \alpha_i &= c_i\varepsilon+{\cal O}\left(\varepsilon^2\right),
    \\
    \beta_1 &=\beta \varepsilon+{\cal O}\left(\varepsilon^2\right),
    \label{eq:beta1 exp}
\end{align}
where $y_0$, $g_2$, $c_i$, and $\beta$ are constants.
Note that the expansion of $\alpha_i$ and $\beta_1$ starts at ${\cal O}(\varepsilon)$,
as modifications of gravity are supposed not to be significant at early times.
As seen below, the background equations
are consistent with Eqs.~\eqref{eq:y0 exp}--\eqref{eq:beta1 exp}.
The expansion coefficients $(y_0, g_2, c_i,\beta)$ are not all independent parameters. We will express some of them in terms of the other coefficients and the parameters of
an underlying model.

Substituting Eqs.~\eqref{eq:y0 exp}--\eqref{eq:beta1 exp}
to Eqs.~\eqref{eq:background phi} and~\eqref{eq:background_p_phi},
one finds, for the attractor solutions (${\cal J}=0$), that
\begin{align}
\rho_\phi &= -\left(g_2+6\beta y_0\right)M^2H^2
\varepsilon+{\cal O}\left(\varepsilon^2\right),\label{rhop1}
\\
p_\phi &=\Bigl[g_2+2
\bigl(c_{\rm B}+3\beta\bigr)y_0-2\beta y^2_0\,
\Bigr]M^2H^2\varepsilon+2M^2\beta y_0 \dot H\varepsilon+{\cal O}\left(\varepsilon^2\right).
\label{pp1}
\end{align}
Noting that $3M^2H^2-\bar\rho_{\rm m}=3M^2H^2\varepsilon= \rho_\phi$,
we have
\begin{align}
g_2=-3\left( 1+2\beta y_0\right).\label{eq47}
\end{align}
The effective dark energy equation of state parameter,
$w_\phi:= p_\phi/\rho_\phi$, can be expanded as
\begin{align}
w_\phi = w^{(0)}+{\cal O}(\varepsilon).\label{eq:w_phi exp}
\end{align}
From Eqs.~\eqref{rhop1}--\eqref{eq47}
and $\dot H/H^2=-3/2+{\cal O}(\varepsilon)$
we obtain
\begin{align}
w^{(0)}=-1 +\frac{2}{3}\left(
c_{\rm B}y_0-\frac{3}{2}\beta y_0-\beta y^2_0
\right).\label{w01}
\end{align}
Using the above expression for $w^{(0)}$, one has the following useful formulas valid up to ${\cal O}(\varepsilon)$:
\begin{align}
\frac{\dot H}{H^2}=-\frac{3}{2}\left(1+w^{(0)}\varepsilon\right)+{\cal O}(\varepsilon^2),
\quad
\frac{\dot\varepsilon}{H}=\left(c_{\rm M}-3w^{(0)}\right)\varepsilon+{\cal O}(\varepsilon^2).
\label{usefulformulas2}
\end{align}

To proceed further, let us assume that $G_2\propto X^p$, where $p$ is a
constant model parameter.
This assumption leads to the relation $XG_{2X} = p G_2$.
Using this assumption and Eq.~\eqref{usefulformulas2}, we find
\begin{align}
0=\dot{\phi}{\cal J}
=\left[
2pg_2+3\bigl(y_0^{-1}c_{\rm M}-2c_{\rm B}\bigr)-9\beta +6\beta\bigl(c_{\rm M}-3w^{(0)}\bigr)+6\beta y_0
\right]M^2H^2\varepsilon +{\cal O}(\varepsilon^2).
\label{jp1}
\end{align}
Equations~\eqref{w01} and~\eqref{jp1} give
\begin{align}
w^{(0)}&=-1-\frac{y_0}{3}\left(c_{\rm H}+2p\right),\label{eq:w0 consistency}
\\
c_{\rm B}&=-p-\frac{c_{\rm H}}{2}+\beta\left(y_0+\frac{3}{2}\right).\label{eq:cB consistency}
\end{align}
Thus, $w^{(0)}$ and $c_{\rm B}$ are expressed in terms of
the model parameter $p$ and the other coefficients.

In the following we consider
tracker solutions characterized by the condition
\begin{align}
H \dot\phi^{2q}={\rm const}\label{trcond}\,,
\end{align}
where $q$ is a constant.
Such tracker solutions
have been studied
in the context of the Horndeski theory~\cite{DeFelice:2011bh, DeFelice:2011aa}
and its
extensions~\cite{Crisostomi:2017pjs, Crisostomi:2018bsp, Frusciante:2018tvu}.
For instance, the cosmological solution discussed in \cite{Crisostomi:2017pjs, Crisostomi:2018bsp}
corresponds to the case with $p=2$ and $q=1$.
In this paper we regard $q$ as another model parameter.
For the solutions satisfying Eq.~\eqref{trcond}, it is easy to see that
\begin{align}
y_0=\frac{3}{4q}.\label{eq:y0 tracker solution}
\end{align}
In what follows we will use $q$ instead of $y_0$

So far we have not imposed $c_{\rm GW}^2=1$ ($\Leftrightarrow \alpha_{\rm T}=0$),
as $\alpha_{\rm T}$ does not appear explicitly in the background equations.
Upon imposing $\alpha_{\rm T}=0$,
it follows from the definitions that $M^2=2G_4$ and $M^2\alpha_{\rm H}=-4XG_{4X}$,
which implies another relation between the parameters:
\begin{align}
\alpha_{\rm M}=- y \alpha_{\rm H}\quad \Rightarrow
\quad
c_{\rm M}=-y_0c_{\rm H}.\label{eq:cM consistency}
\end{align}
Thus, under the assumption of $c_{\rm GW}^2=1$,
we have four independent parameters,
$(p,q, c_{\rm H}, \beta)$, in terms of which
$g_2$, $c_{\rm M}$, $c_{\rm B}$, as well as $w^{(0)}$,
can be expressed.

\section{Constraining DHOST cosmology}

\subsection{Growth index}
\label{sec:linear}

Let us derive the solution to~\eqref{eq:evolution f} in a series expansion form in terms of $\varepsilon$.
We start with expanding $\varsigma$ and $\Xi_\Phi$ in terms of $\varepsilon$.
Since $\alpha_i={\cal O}(\varepsilon)$ and $\beta_1={\cal O}(\varepsilon )$, we have
$\varsigma\to 0$ and $\Xi_\Phi\to 1$ for $\varepsilon\to 0$,
so that, to ${\cal O}(\varepsilon)$, $\varsigma$ and $\Xi_\Phi$ can be
written as
\begin{align}
\varsigma= \varsigma^{(1)}\varepsilon+{\cal O}(\varepsilon^2),
\quad
\Xi_\Phi=1+\Xi_\Phi^{(1)}\varepsilon+{\cal O}(\varepsilon^2),
\label{vasigma1Xi1}
\end{align}
where
$\varsigma^{(1)}$ and $\Xi_\Phi^{(1)}$
can be written in terms of
the parameters introduced in the previous section.
See Appendix~\ref{sec:munukappa} for their explicit expressions.
Then, Eq.~\eqref{eq:evolution f} reduces to
\begin{align}
\bigl(c_{\rm M}-3w^{(0)}\bigr)\varepsilon\frac{df}{d\varepsilon}
+\left[\frac{1}{2}+\left(\varsigma^{(1)}-\frac{3}{2}w^{(0)}\right)\varepsilon\right]f
+f^2-\frac{3}{2}\biggl[1-\left(1-\Xi_\Phi^{(1)}\right)\varepsilon\biggr]
+ {\cal O}(\varepsilon^2)=0,
\end{align}
where we used Eq.~\eqref{usefulformulas2}.
The solution to this equation is given by
\begin{align}
f=1-  \left[
\frac{3\bigl(1-w^{(0)}\bigr)+2\varsigma^{(1)}-3\Xi_\Phi^{(1)}}{5-6w^{(0)}+2c_{\rm M}}\right]\varepsilon+
{\cal O}(\varepsilon^2).\label{solf1}
\end{align}
From the solution~\eqref{solf1} we immediately obtain
\begin{align}
\gamma =
\frac{3(1-w^{(0)})+2\varsigma^{(1)}-3\Xi_\Phi^{(1)}}{5-6w^{(0)}+2c_{\rm M}} + {\cal O}(\varepsilon).
\label{gamma_para}
\end{align}
It is easy to see that the standard result $\gamma=6/11$
is recovered for $w^{(0)}=-1$, $c_{\rm M}=\varsigma^{(1)}=\Xi_\Phi^{(1)}=0$.
Substituting the explicit expressions for $\Xi_\Phi^{(1)}$ and $\varsigma^{(1)}$ [Eqs.~\eqref{eq:Xi_Phi1} and \eqref{eq:varsigma1}]
into Eq.~\eqref{gamma_para}, one can evaluate
an approximate form of the growth index $\gamma$
during the matter dominated era and the early stage of the dark energy
dominated era satisfying $\varepsilon \ll 1$:
\begin{align}
    \gamma
        =&\frac{3\bigl[\bigl(1-w^{(0)}\bigr) -c_{\rm T}\bigr]}{5-6w^{(0)}+2c_{\rm M}}
                    -\frac{2}{\Sigma}\Bigl[ c_{\rm B}-c_{\rm M}+c_{\rm T}-\beta\bigl( c_{\rm M}-3w^{(0)}\bigr)\Bigr]^2
    \notag\\
    &
        +\frac{c_{\rm H}+\beta}{\Sigma}
            \biggl\{
            6\bigl( 1+w^{(0)}\bigr) +\bigl(c_{\rm M}-c_{\rm T}\bigr)\Bigl[1-2\bigl( c_{\rm M}-3w^{(0)}\bigr)\Bigr]
    \notag\\
    &\quad\quad
            +\Bigl[ c_{\rm B}-\beta\bigl( c_{\rm M}-3w^{(0)}\bigr) \Bigr]\Bigl[5-2\bigl( c_{\rm M}-3w^{(0)}\bigr)\Bigr]
            +5\bigl(c_{\rm H}+\beta\bigr)\bigl( c_{\rm M}-3w^{(0)}\bigr)
            \biggr\} + {\cal O}(\varepsilon)
    \,,\label{eq:gamma}
\end{align}
where
\begin{align}
     \Sigma =\frac{1}{3}\bigl(5-6w^{(0)}+2c_{\rm M}\bigr)
        \biggl\{3\bigl(1 +w^{(0)}\bigr) +2\left( c_{\rm M}-c_{\rm T}\right) +\Bigl[1 -2\bigl(c_{\rm M} -3w^{(0)}\bigr)\Bigr]\Bigl[c_{\rm B} -c_{\rm H}-\beta\bigl(c_{\rm M}-3 w^{(0)}+1\bigr)\Bigr]\biggr\}
    \,.\label{eq:Sigma}
\end{align}
The first two terms in Eq.~\eqref{eq:gamma} are the generalization of the previous results derived in the case of the Horndeski theory~\cite{Yamauchi:2017ibz}
and the third term appears when
at least either of $c_{\rm H}$ and $\beta$ is nonvanishing,
namely when one considers theories beyond Horndeski.
Equation~\eqref{eq:gamma} is general in the sense that
we have not yet imposed $\alpha_{\rm T}=0$.
Now, imposing $\alpha_{\rm T}=0$ ($\Rightarrow c_{\rm T}=0$),
as discussed around Eq.~(\ref{eq:cM consistency}), $\gamma$ can be written
in terms of $(p,q,c_{\rm H},\beta)$ as
\begin{align}
    \gamma
        =&\frac{3}{2(-3+6p+10q)(3p+11q)}
            \biggl\{
                \Bigl[\left( p+4q\right)\left( -3+6p+10q\right) -8pq^2\Bigr]
    \notag\\
    &\quad
                +\frac{1}{2}\Bigl[ \left( -3+6p+10q\right)+8q\left(3p+2q\right)\Bigr] c_{\rm H}
                +\frac{3q(1+2q)(-3+6p+16q)(c_{\rm H}+\beta)^2}{2pq+3qc_{\rm H}+(3p+5q)\beta}
            \biggr\}
    \,.\label{eq:gamma pq}
\end{align}

\subsection{Observational constraints}
\label{sec:constraint}

\begin{figure}[tb]
    \includegraphics[keepaspectratio=true,height=70mm]{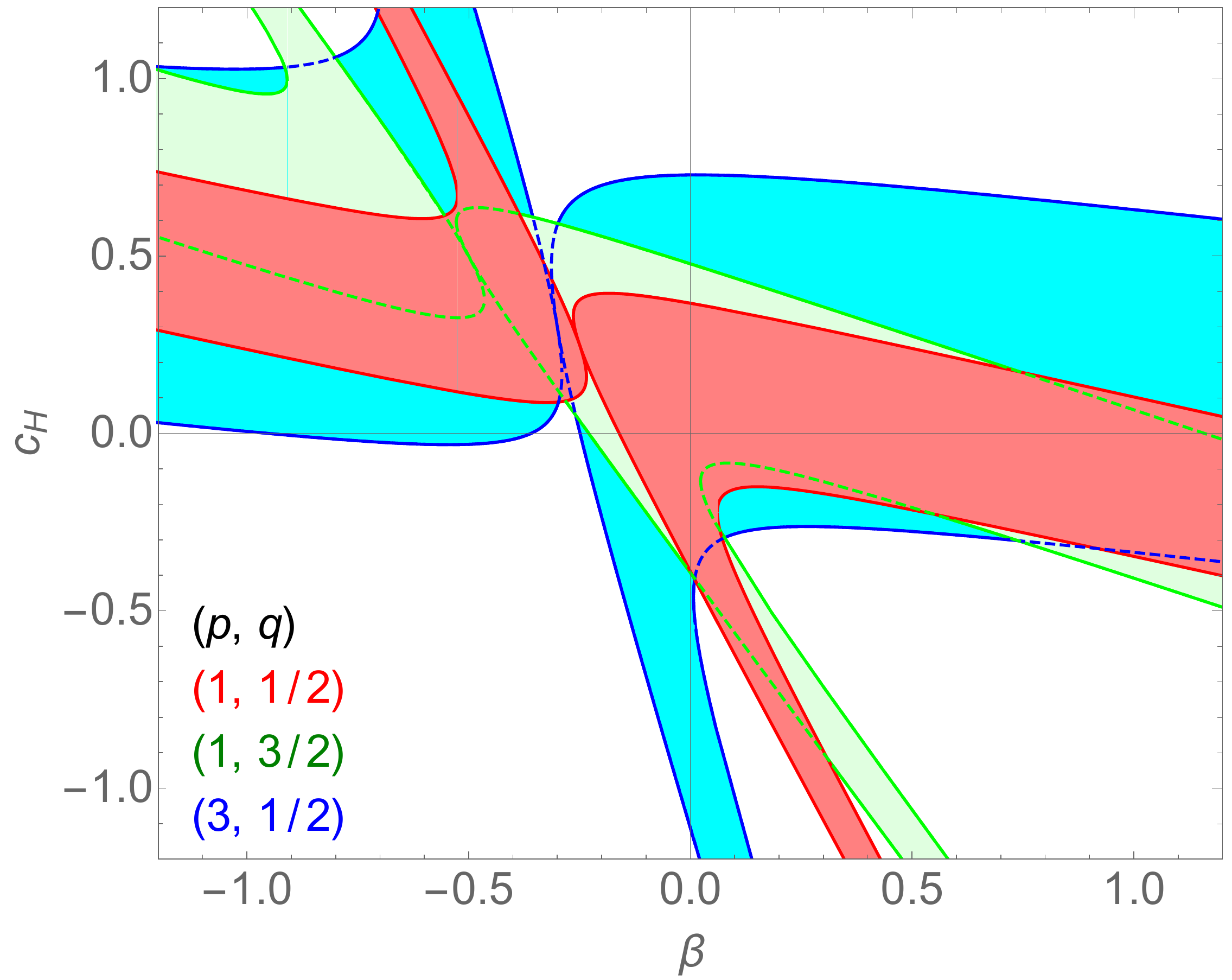}
     \caption{
        The allowed parameter region in the $\beta$-$c_{\rm H}$ plane
    obtained from the gravitational growth index in the shift-symmetric
     quadratic DHOST cosmology after GW170817.
The parameters are given by $(p,q)=(1,1/2)$ (red), $(1,3)$ (green),
and $(3,1/2)$ (blue). The first set of the parameters corresponds
to the solution discussed in~\cite{Crisostomi:2017pjs, Crisostomi:2018bsp}.
    }
     \label{fig:1}
\end{figure}

In this section,
we investigate constraints on
DHOST theories based on current observational limits on
the gravitational growth index $\gamma$.
For instance, clustering measurements from
the BOSS DR12 give the limit as $\gamma=0.52\pm 0.10$ in Ref.~\cite{Grieb:2016uuo} (based on the analysis in Fourier space)
and $\gamma =0.609\pm 0.079$ in Ref.~\cite{Sanchez:2016sas} (based on the analysis in configuration space).
The constraints from BOSS DR14
are given as $\gamma=0.55\pm 0.19$ in Ref.~\cite{Gil-Marin:2018cgo}
and $\gamma=0.580\pm 0.082$ in Ref.~\cite{Zhao:2018jxv} (by adding tomographic analysis).
Since the typical value of the deviation from the central value of $\gamma$ in the current observations
as shown above can be roughly estimated as $\lesssim{\cal O}(0.1)$, let us employ $\gamma =6/11\pm 0.1$ as a conservative constraint.
For a given set of the model parameters $(p,q)$, this can be translated into
constraints on $(\beta, c_{\rm H})$ using Eq.~\eqref{eq:gamma pq}.
The parameter regions in the $\beta$-$c_{\rm H}$ plane
allowed by the constraint $\gamma =6/11\pm 0.1$ are plotted in Fig.~\ref{fig:1}
for $(p,q)=(1,1/2)$ (red), $(1,3/2)$ (green), and $ (3,1/2)$ (blue).
One finds from Fig.~\ref{fig:1}
that a constant-$\gamma$ curve for fixed $p$ and $q$
is a hyperbola in the $\beta$-$c_{\rm H}$ plane
for $(p,q)$ and $\gamma$ that we are considering. This means that
we have degeneracy between $c_{\rm H}$ and $\beta$
in the observations of the growth index.
In contrast, in the GLPV theory we have $\beta=0$,
and hence we can obtain for instance the following constraints on $c_{\rm H}$:
$-0.4\leq c_{\rm H}\leq 0.4$ for $(p,q)=(1\,,1/2)$,
$-0.4\leq c_{\rm H}\leq 0.5$ for $(1\,,3/2)$, and
$-1.1\leq c_{\rm H}\leq 0.7$ for (3\,,1/2).
Deriving the constraints for other values of $(p,q)$ is straightforward.
It should be emphasized that the constraints we have obtained in Fig.~\ref{fig:1}
are those at high redshifts satisfying $\Omega_{\rm m}\simeq 1$.

To compare our results with previously known constraints, it is necessary
to make further assumptions that connect the series expansion
of $\alpha_{\rm H}$ and $\beta_1$ to their present values.
Specifically, we assume that
$\alpha_{\rm H}=c_{\rm H}\left( 1-\Omega_{\rm m}\right)$,
$\beta_1=\beta\left( 1-\Omega_{\rm m}\right)$, and the leading order expression of $\gamma$ [Eq.~\eqref{eq:gamma pq}]
are valid all the way up to the present time.
Hereafter we focus on the specific parameter values
$(p,q)=(1,1/2)$, which corresponds
to the model discussed in~\cite{Crisostomi:2017pjs, Crisostomi:2018bsp},
and demonstrate the allowed parameter region.
Though details of constraints will be different for
different choices of $(p,q)$, we expect that the order of the bounds is approximately the same.

Existing constraints on DHOST theories mainly come from
the Newtonian stellar structure modified due to
the partial breaking of the Vainshtein mechanism,
which is characterized by a single parameter
$\Upsilon_1%
:=-2(\alpha_{\rm H} +\beta_1)^2/(\alpha_{\rm H} +2\beta_1)$
(the definition here is for theories with
$c_{\rm GW}^2=1$)~\cite{Kobayashi:2014ida, Langlois:2017dyl, Dima:2017pwp}.
The lower bound on
$\Upsilon_1$ has been obtained
from the requirement that gravity is attractive
at the stellar center: $\Upsilon_1>-2/3$~\cite{Saito:2015fza}.
The upper bound is given by
comparing the minimum mass of stars with the hydrogen burning
with the minimum mass of observed red dwarfs:
$\Upsilon_1<1.6$~\cite{Sakstein:2015aac}.

There are several attempts for improving the above
bounds~\cite{Jain:2015edg,Babichev:2016jom,Saltas:2018mxc},
including the one concerning
the speed of sound in the atmosphere of the Earth~\cite{Babichev:2018rfj}.
Aside from the constraints from the Newtonian stellar structure,
another constraint has been proposed,
which comes from precise observations of the Hulse-Taylor pulsar.
This can severely constrain the effective parameters
through the coupling of gravitational waves to matter~\cite{Jimenez:2015bwa, Dima:2017pwp}
:$-7.5\times 10^{-3}\leq\alpha_{\rm H}+3\beta_1\leq 2.5\times 10^{-3}$.
However, when deriving this result,
several assumptions have been made and the resultant constraint
would depend on the details of how the screening mechanism operates in
a binary system.
In this paper, we try to constrain the effective parameters
without taking into account these potentially more stringent bounds, and
use the most conservative constraint: $-2/3 < \Upsilon_1 < 1.6$.

\begin{figure}[tb]
    \includegraphics[keepaspectratio=true,height=70mm]{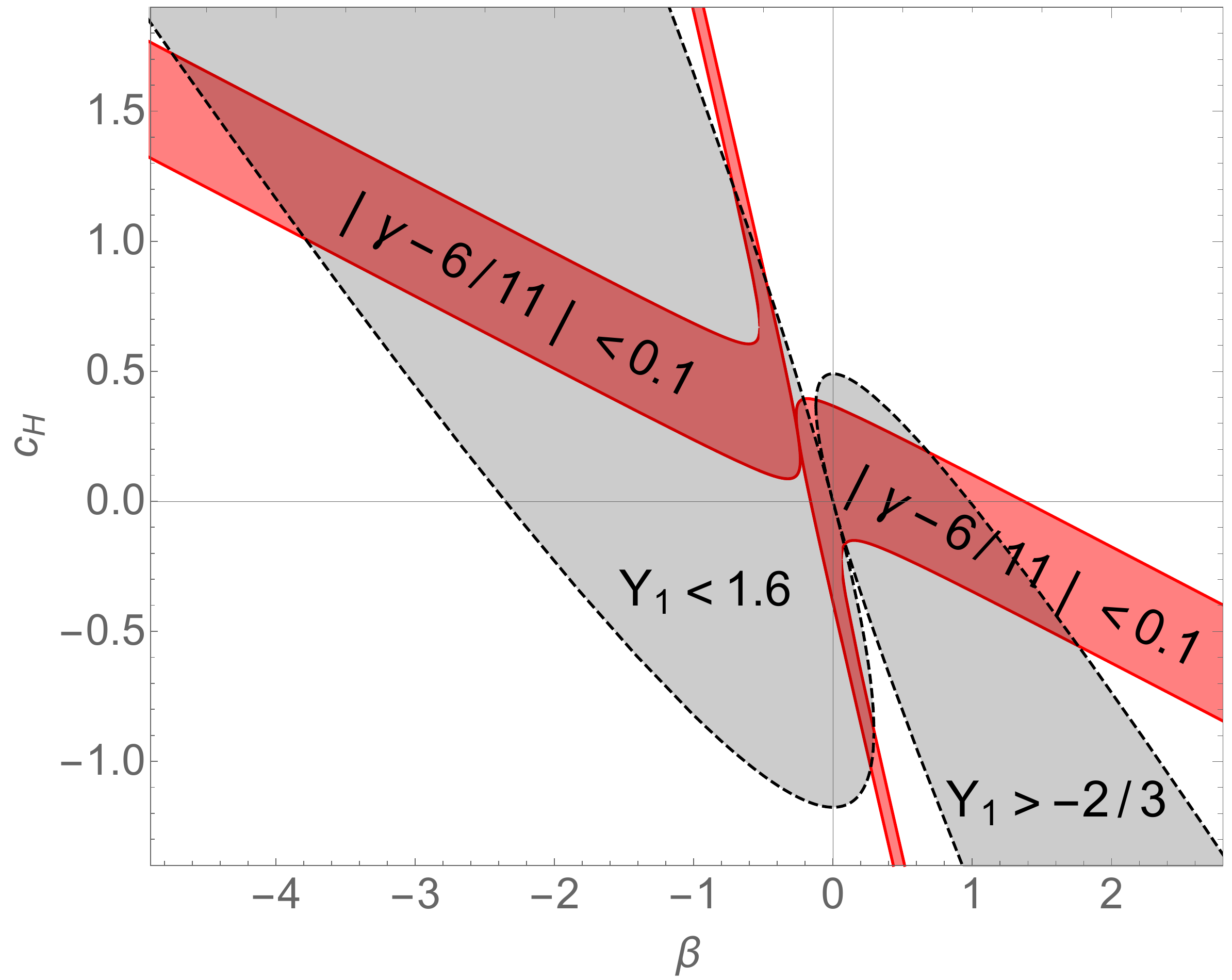}
     \caption{
        The allowed parameter region obtained from
    the gravitational growth index in the shift-symmetric quadratic DHOST cosmology
after GW170817 is shown by the red area
in the $\beta$-$c_{\rm H}$
plane. The parameters are given by
    $p=1,q=1/2$.
        For comparison, the existing constraints
        $-2/3\leq\Upsilon_1\leq 1.6$~\cite{Saito:2015fza, Sakstein:2015aac}
are shown by the gray area.
    }
     \label{fig:2}
\end{figure}

We plot in Fig.~\ref{fig:2} the allowed parameter region in the
$\beta$-$c_{\rm H}$ plane obtained from
the constraints on the growth index (red) and stellar structure (black).
As shown in Fig.~\ref{fig:2}, combining our results and the conservative constraints discussed above can break the degeneracy between
$c_{\rm H}$ and $\beta$ without using the Halse-Taylor pulsar bound.
The overlap region between these gives the constraints on both parameters:
$-1.0\leq  c_{\rm H}\leq 1.7$ and $-4.7\leq\beta\leq 1.8$.

Note that recently it was pointed out in Ref.~\cite{Creminelli:2018xsv}
that the absence of gravitational wave decay into scalar modes
requires $\alpha_{\rm H} +2\beta_1=0$. As seen from the fact that 
the denominator of $\Upsilon_1$ vanishes when this is satisfied,
this is a special case which has not been explored so far. 
It would be interesting to  investigate the behavior of
gravity in this limiting case in detail, but it is beyond the scope of this paper, and we do not consider
this constraint.

\section{Summary}
\label{sec:discussion}

In this paper, we have considered a possibility to constrain
degenerate higher-order scalar-tensor (DHOST) theories
by using the information about the linear growth of matter density fluctuations.
In DHOST theories, the evolution equation for the linear matter density fluctuations is modified in such a way that the effective gravitational coupling is changed by the factor $\Xi_\Phi$ and the friction term has an additional contribution $\varsigma H$,
both of which can be expressed in terms of
the effective parameters $\alpha_i$ and $\beta_i$ used in the literature.

We have constructed cosmological models
in DHOST theories as
a series expansion in terms of $1-\Omega_{\rm m}$.
In doing so, we have assumed for simplicity that cosmological solutions under consideration are attractors in shift-symmetric theories and subject to the tracker ansatz. The resultant cosmology
is characterized by two model parameters $(p,q)$ and four independent
effective parameters in general (i.e., six parameters in total),
and upon imposing $c_{\rm GW}^2=1$ 
the number of independent parameters reduces to four in total.
Our construction thus provides a concise description
of DHOST cosmology during the matter dominated era and
the early stage of the dark energy dominated era.

We have then explicitly expressed the gravitational growth index $\gamma$ in terms of $(p,q)$ and the effective parameters.
We have found that
the constant-$\gamma$ curve in the $\beta$-$c_{\rm H}$ plane generically
is a hyperbola for $c_{\rm GW}^2=1$ and fixed $(p,q)$.
One can thus obtain constraints on a certain combination
of the effective parameters at high redshifts
by using the observations of the growth index alone.

Under the additional assumption that
our leading order results in $1-\Omega_{\rm m}$ expansion
can be extrapolated all the way
to the present time,
we have compared the constraints from the growth index 
with the previously known bounds.
Combining our results and
the constraints from modifications of the gravitational law inside stellar objects,
we have shown that the parameter degeneracy between $\alpha_{\rm H}/(1-\Omega_{\rm m})$ and $\beta_1/(1-\Omega_{\rm m})$ could
be broken without using the Hulse-Taylor pulsar constraint,
though our results slightly depend on the model parameters.
Future-planned observations for large-scale structure would exclude
the currently allowed region of the parameter space and serve as tests of the viability of DHOST theories.

\acknowledgements

We would like to thank
Rampei Kimura and Kazuhiro Yamamoto for fruitful discussion.
This work was supported in part by the
JSPS Research Fellowships for Young Scientists
No.~17J04865 (S.H.),
the JSPS Grants-in-Aid for Scientific Research
 Nos.~16H01102, 16K17707 (T.K.), 15K17659 (S.Y.), 17K14304 (D.Y.),
 MEXT-Supported Program for the Strategic Research Foundation at Private Universities,
 2014-2017 (S1411024) (T.K. and S.Y.), and MEXT KAKENHI Grant Nos.~15H05888 (T.K. and S.Y.), ~17H06359 (T.K.) and ~18H04356 (S.Y.).

\appendix


\section{Explicit expressions for some coefficients}
\label{sec:munukappa}

Let us write down explicitly
the coefficients in Eqs.~\eqref{eq:Q=delta}--\eqref{eq:Phi=delta}.
The coefficients in Eq.~\eqref{eq:Q=delta} are given by
\begin{align}
    &\nu_Q=\frac{3\Omega_{\rm m}}{2{\cal Z}}{\cal N}_\Phi
    \,,\label{eq:nu_Q}\\
    &\kappa_Q=\frac{3\Omega_{\rm m}}{8{\cal Z}}
            \biggl\{
                \Bigl[ c_1+2\bigl(2\beta_1+\beta_3\bigr)\Bigr]{\cal F}_\Phi
                +\bigl( c_2-4\alpha_{\rm H}\bigr){\cal F}_\Psi
    \notag\\
    &\quad\quad\quad\quad\quad
                -\frac{M^2}{H}\biggl[
                    2\bigl(2\beta_1+\beta_3\bigr)\frac{\D}{\D t}\left(\frac{{\cal F}_\Phi}{M^2}\right)
                    -4\alpha_{\rm H}\frac{\D}{\D t}\left(\frac{{\cal F}_\Psi}{M^2}\right)
                    \biggr]
            \biggr\}
    \,,\label{eq:kappa_Q}
\end{align}
where we have defined the some dimensionless parameters as
\begin{align}
    &{\cal S}{\cal N}_\Phi =\alpha_{\rm H}\bigl( 1+\alpha_{\rm H}\bigr) -\frac{1}{2}\bigl( 1+\alpha_{\rm T}\bigr)\bigl( 2\beta_1+\beta_3\bigr)
    \,,\\
    &{\cal S}{\cal F}_\Phi =1+\alpha_{\rm T}
    \,,\ \ \
    {\cal S}{\cal F}_\Psi =1+\alpha_{\rm H}
    \,,\\
    &{\cal S}=\bigl( 1+\alpha_{\rm H}\bigr)^2-\frac{1}{2}\bigl( 1+\alpha_{\rm T}\bigr)\beta_3
    \,,
\end{align}
The denominator ${\cal Z}$ can be written as
\begin{align}
    &{\cal Z}
        =\frac{1}{4}
            \Biggl\{
                {\cal E}_\Phi c_1+{\cal E}_\Psi c_2-\biggl[
                 b_3+\frac{2(2\beta_1+\beta_3)}{H}\dot{\cal E}_\Phi-\frac{4\alpha_{\rm H}}{H}\dot{\cal E}_\Psi
                \biggr]
            \Biggr\}
\end{align}
where $c_1$, $c_2$, and $b_3$ were defined
in Eqs~\eqref{eq:c1}, \eqref{eq:c2}, and \eqref{eq:b3}. We have also defined
\begin{align}
    &{\cal S}{\cal E}_\Phi=b_1\bigl( 1+\alpha_{\rm T}\bigr) -b_2\bigl( 1+\alpha_{\rm H}\bigr)
    \,,\\
    &{\cal S}{\cal E}_\Psi=b_1\bigl( 1+\alpha_{\rm H}\bigr) -\frac{1}{2}b_2\beta_3
    \,.
\end{align}
The coefficients in
Eqs.~\eqref{eq:Psi=delta} and \eqref{eq:Psi=delta} are
\begin{align}
    &\mu_a ={\cal N}_a\nu_Q
    \,,\label{eq:mu_a}\\
    &\nu_a =-{\cal E}_a\nu_Q +{\cal N}_a\left[\kappa_Q+\frac{1}{a^2H^2}\frac{\D}{\D t}\left(a^2H\nu_Q\right)\right]
    \,,\\
    &\kappa_a =  \frac{3}{2}\Omega_{\rm m}{\cal F}_a -{\cal E}_a\kappa_Q +\frac{{\cal N}_a}{a^2H^3}\frac{\D}{\D t}\left(a^2H^2\kappa_Q\right)
    \,,\label{eq:kappa_a}
\end{align}
for $a=\Psi\,,\Phi$, where
\begin{align}
    {\cal S}{\cal N}_\Psi =-\bigl( 1+\alpha_{\rm H}\bigr)\beta_1 -\frac{1}{2}\beta_3
    \,.
\end{align}
The coefficients in Eq.~\eqref{vasigma1Xi1} are given by
\begin{align}
    &\varsigma^{(1)} = \frac{3}{Z}\left(c_{\rm H} +\beta\right)^2\bigl(c_{\rm M} -3w^{(0)}\bigr)\label{eq:varsigma1}
    \,,\\
    &\Xi^{(1)}_{\Phi}
    =c_{\rm T}+\frac{2}{Z}\Bigl[ c_{\rm B}-c_{\rm M}+c_{\rm T}-\beta\bigl( c_{\rm M}-3w^{(0)}\bigr)\Bigr]^2
    \notag\\
    &\quad\quad\quad
        -\frac{c_{\rm H}+\beta}{Z}
            \biggl\{
            6\bigl( 1+w^{(0)}\bigr) +\bigl(c_{\rm M}-c_{\rm T}\bigr)\Bigl[ 1-2\bigl( c_{\rm M}-3w^{(0)}\bigr)\Bigr]
    \notag\\
    &\quad\quad\quad\quad\quad\quad
            +\Bigl[ c_{\rm B}-\beta\bigl( c_{\rm M}-3w^{(0)}\bigr) \Bigr]\Bigl[ 5-2\bigl( c_{\rm M}-3w^{(0)}\bigr)\Bigr]
            +3\bigl(c_{\rm H}+\beta\bigr)\bigl( c_{\rm M}-3w^{(0)}\bigr)
            \biggr\}\,, \label{eq:Xi_Phi1}
\end{align}
where
\begin{align}
     Z =3\bigl(1 +w^{(0)}\bigr) +2c_{\rm M} +\Bigl[1 -2\bigl(c_{\rm M} -3w^{(0)}\bigr)\Bigr]\Bigl[c_{\rm B} -c_{\rm H}-\beta\bigl(c_{\rm M}-3 w^{(0)}+1\bigr)\Bigr]
    \,.
\end{align}
We can then finally obtain the explicit expression $\gamma$ in the main text by sustituting Eqs.~\eqref{eq:varsigma1} and \eqref{eq:Xi_Phi1} into Eq.~\eqref{gamma_para}.


\end{document}